\documentclass[aps, prd, amsmath, amssymb, reprint]{revtex4-2}

\usepackage[colorlinks = true,
            linkcolor = blue,
            urlcolor  = blue,
            citecolor = blue, 
            anchorcolor = blue]{hyperref}
 
\usepackage{graphicx}
\usepackage{amsmath}
\usepackage{booktabs}
\usepackage{accents}
\usepackage{multirow}
\newcommand\brobor{\smash[b]{\raisebox{0.6\height}{\scalebox{0.5}{\tiny(}}{\mkern-1.5mu\scriptstyle-\mkern-1.5mu}\raisebox{0.6\height}{\scalebox{0.5}{\tiny)}}}}

\begin{document} 

\title{Indirect search of Heavy Neutral Leptons using the DUNE Near Detector}
\author{S. Carbajal} 
\affiliation{Secci\'on F\'isica, Departamento de Ciencias, Pontificia Universidad Cat\'olica del Per\'u, Apartado 1761, Lima, Per\'u}
\author{A.M. Gago}
\affiliation{Secci\'on F\'isica, Departamento de Ciencias, Pontificia Universidad Cat\'olica del Per\'u, Apartado 1761, Lima, Per\'u}  

\begin{abstract}
We evaluate the potential of the DUNE Near Detector (DUNEND) for establishing bounds for heavy neutral leptons (HNL). This is achieved by studying how the presence of HNLs affects the production rates of active neutrinos, therefore, creating a deficit in the neutrino charged current (CC) events at the LArTPC of the DUNEND. The estimated bounds on HNLs are calculated for masses between 1 eV and 500 MeV. We consider 10 years of operation (5 in neutrino and antineutrino mode) and obtain limits of $|U_{\mu4}|^2 < 9 \times 10^{-3} (4 \times10^{-2})$ and $|U_{e4}|^2 < 7\times10^{-3} (3 \times10^{-2})$ for masses below 10 MeV and a 5\%(20\%) overall normalization uncertainty in the neutrino charged current event rates prediction. These limits, within the region of masses below 2(10) MeV, are better than those that can be achieved by DUNE direct searches for the case of a 5\%(20\%) uncertainty. When a conservative 20\% uncertainty is present, our limits can only improve current constraints on $|U_{e4}|^2$ by up to a factor of 3 in a small region around 5 eV and set limits on $|U_{\mu4}|^2$ in a mass region free of constraints (40 eV - 1 MeV).
\end{abstract}

\maketitle


\section{Introduction}

Heavy neutral leptons (HNLs) are singlet (right-handed) fermions states introduced for explaining the non-zero neutrino masses, interacting, via Yukawa coupling, with the Higgs boson and the leptonic doublet, a Dirac mass term, and also appearing in the Majorana mass term. The nearly sterile states that arise after the diagonalization of the mass terms mentioned before interact with matter via suppressed mixing to the active neutrinos of the Standard Model (SM) \cite{Drewes:2013gca,Atre:2009rg}. 

The HNLs are candidates to solve important particle physics and cosmology issues~\cite{Drewes:2013gca}. They can help explain the smallness of the active neutrino masses via the Seesaw mechanism \cite{King:2003jb}, act as possible dark matter candidates~\cite{Drewes:2016upu} and also explain the baryon asymmetry of the universe through their role in leptogenesis (see \cite{Drewes:2013gca,Davidson:2008bu} and references therein). On the other hand, neutrino oscillations involving light sterile states have been proposed to explain the excess of electron antineutrino and neutrino events at LSND and MiniBoone, respectively, as well as the deficit of electron antineutrino events at reactor experiments~\cite{Abazajian:2012ys}. The HNL masses required for solving the before mentioned problems fall within a mass range that spans from 
keV to TeV. As a consequence of their relevance, there have been several HNL searches in this wide mass range, placing limits on the possible values of the HNL mass $m_N$ and its mixing to the SM neutrinos $|U_{\alpha4}|^2$ \cite{Atre:2009rg,Berryman:2019dme}.  

In particular, searches of HNLs in the range of masses of 1-400 MeV have been conducted in accelerator-based experiments through searches for low-energy peaks in the energy spectrum of the muons resulting from pion ($\pi^\pm \rightarrow \mu^\pm \nu_H$) and kaon decays ($K^\pm \rightarrow \mu^\pm \nu_H$) \cite{Shrock:1980vy,Shrock:1980ct,PIENU:2019usb,PhysRevD.91.052001}. With no positive results found so far, they obtain upper bounds for  $|U_{\mu4}|^2$  such as $10^{-6}$ for $m_N\sim 10$ MeV and $10^{-9}$ for $m_N\sim300$ MeV.

This work aims to assess the sensitivity of the DUNE experiment in setting upper limits for $|U_{\mu4}|^2$ and $|U_{e4}|^2$ for masses below 500 MeV. We achieved the latter by comparing the energy distributions of the neutrino CC event rates with and without HNLs at the DUNE Near Detector (DUNEND) \cite{DUNE:2020lwj}. We find that the presence of HNLs creates a deficit of CC events that is not generated by neutrino oscillations, but instead by the combination of kinematic effects in the production and decay chains of HNLs: the decrease of the branching ratios of active neutrino production, the large lifetimes of HNLs and the fact that active neutrinos born from HNLs have angular distributions spanned outside the detector coverage. We consider the decrease in CC events as an indirect signal of HNLs and use it to set limits on the mixing parameters. Additionally, we present an analysis of the possibility of finding confidence regions for the values of $(m_N, |U_{\alpha4}|^2)$ if a deficit of CC events is found at DUNE \cite{DUNE:2020ypp}.\\

This paper goes as follows: in the second section we
discuss the theoretical framework of HNL production and decay. Then, in the third one, we describe the experimental setup. In the fourth section, the details of our simulation are given. While, in the fifth one, our results are presented. We draw our conclusions in the final section. 


\section{Theoretical Framework}
\label{theory}

As we already mentioned, the nearly sterile mass eigenstates couple to the active flavor states via an extended version of the Pontecorvo-Maki-Nakagawa matrix (PMNS) \cite{Giganti:2017fhf}, which can be expressed as follows:
\begin{align}
\nu_\alpha = \sum_{i=1,2,3}U_{\alpha i}\nu_i + U_{\alpha 4}N,
\end{align}
where $N$ represents the HNL field. It is also helpful to write the new active neutrino flavor states in terms of the flavor states of the SM $\nu_\alpha^{\text{SM}}$, which represent the neutrino flavor states when the values of the $3 \times 3$ PMNS mixing matrix are assumed. This can be done by the approximation \cite{Gronau:1984ct}
\begin{align}
\nu_\alpha \approx \nu_\alpha^{\text{SM}} \left(1 - \frac{|U_{\alpha4}|^2}{2}\right) + U_{\alpha 4} N.
\end{align}

Due to the connection above, the HNLs can be produced in any weak decay involving active neutrinos. The production rate of HNLs depends kinematically on its mass $m_N$, the strength of its mixing to active neutrinos $|U_{\alpha4}|^2$ and the nature of the decaying particle that produces it, which, from now on, we will refer to as its parent. In this work, we are interested in HNLs with masses below the kaon mass $(m_K)$. The production of HNLs from kaon and pion decays, followed by the muon decays, dominate at the typical energies of beam dump experiments such as DUNE. Their production from heavier particles, such as $D$ mesons or $\tau$ leptons, is also possible, but it is rare since the production of the latter is heavily suppressed in comparison to the light mesons. Table \ref{tab:prodchannels} shows the dominant HNL production channels from light leptons and mesons, along with the maximum kinematically allowed values of the masses for the HNLs. A rough estimation of these values is obtained by subtracting the total rest mass of the particles produced, other than the HNLs, from the corresponding mass of their parent particle.

\begin{table}[!h]
\caption{\label{tab:prodchannels}Channels considered for the production of HNLs. The maximum possible value of $m_N$ is shown for each channel. Charged conjugate channels were also considered.}
\centering
  \begin{tabular}{llr}\toprule  
              &   Channel                 &   $m_N$(MeV)	\\ \midrule
  $\mu^+ \to$ & $e^+\nu_e\bar{\nu}_\mu$   & 105.14      	\\ 
  $\pi^+ \to$ & $\mu^+\nu_\mu$            & 33.91       	\\ 
              & $e^+\nu_e$                & 139.06      	\\ 
  $K_L^0 \to$ & $\pi^\pm e^\mp \nu_e$     & 357.12      	\\ 
              & $\pi^\pm \mu^\mp \nu_\mu$ & 252.38       	\\ 
	$K^+ \to$ & $\mu^+\nu_\mu$        & 387.81    \\
	& $\pi^0 e^+ \nu_e$     & 358.19    \\ 
	& $\pi^0 \mu^+ \nu_\mu$ & 253.04    \\ 
	& $e^+\nu_e$            & 493.17    \\ \bottomrule
  \end{tabular}
\end{table}

\begin{figure}[!h]
	\centering
	\includegraphics[width=0.9\linewidth]{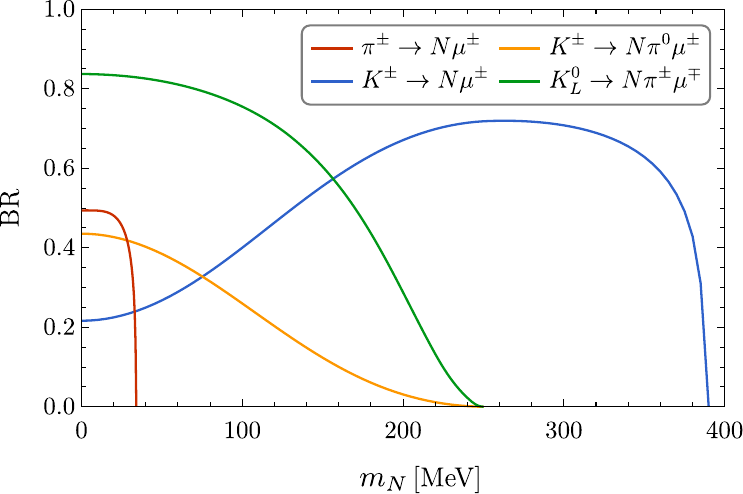}
  \caption{\label{fig:br2hnlLight}
    Branching ratios of the dominant HNL production channels  for $|U_{\mu4}|^2=1$.
  }
\end{figure}

We calculated the branching ratios for HNL production by using the formulas from Ref. \cite{Bondarenko:2018ptm}. For instance, Fig. \ref{fig:br2hnlLight} shows the branching ratios of the dominant HNL production channels below the kaon mass for $|U_{\mu 4}|^2=1$. We can note that almost all the branching ratios decrease with $m_N$, with the only exception being the leptonic decays of charged kaons, $K^\pm\rightarrow N\mu^\pm$. Above 34 MeV, the production from pions is kinematically forbidden; this is important since this means that all heavy neutral leptons above this mass will be produced only from kaon decays. As the value of $m_N$ increases, the branching ratio of $K^\pm\rightarrow N\mu^\pm$ keeps increasing as well, surpassing the branching ratios of $K^\pm\rightarrow N\pi^0\mu^\pm$ at around 80 MeV and of $K_L^0\rightarrow N\pi^0\mu^\pm$ at around 160 MeV. Finally, the branching ratio of $K^\pm\rightarrow N\mu^\pm$ reaches its maximum at around 260 MeV and then decreases until it is kinematically forbidden. The endpoint of each branching ratio corresponds to the maximum $m_N$ given in~Table \ref{tab:prodchannels}.  

The production of HNLs via semileptonic decays involves hadronic currents that cannot be calculated from first principles due to the non-perturbative nature of QCD at low energies. Therefore, the dynamics of these decays are modeled by form factors that represent the momentum distribution of the quarks inside the mesons and parametrize the momentum transfer between the hadronic current and the lepton pair \cite{Richman:1995wm}. For all the semileptonic decays in Table \ref{tab:prodchannels}, we used the form factors presented in \cite{Bondarenko:2018ptm}. \\

After their production, all the HNLs propagate and then decay on flight via mixing with active neutrinos. Table \ref{tab:hnldecays} shows all the decay channels for the HNLs considered in this work. We included all the kinematically allowed decays to final states involving pseudoscalar mesons as well as pure leptonic decays for $m_N<m_K$. A more complete table can be found in \cite{Ballett:2019bgd}.

\begin{table}[!h]
	\caption{ \label{tab:hnldecays}
		HNL decay channels considered in this work. The minimum required value of $m_N$ is shown for each channel.
	}
	\centering
	  \begin{tabular}{cc}\toprule
			\multirow{2}{*}{Channel}	& Threshold		\\ 
																& [MeV]				\\ \midrule
			$\nu\nu\nu$								&	$10^{-9}$		\\
			$\nu e^+e^-$ 							&	1.02				\\
			$\nu e^\pm \mu^\mp$ 			&	106.17			\\
			$\nu \pi^0$								&	134.98			\\
			$e^\mp \pi^\pm$						&	140.08			\\
			$\nu \mu^+\mu^-$					&	211.32			\\
			$\mu^\mp \pi^\pm$ 			&	245.23		\\ \bottomrule
		\end{tabular}
\end{table}

The partial width of a HNL decay channel involving a final lepton $l_\alpha$ or light neutrino $\nu_\alpha$ is directly proportional to the mixing parameter squared $|U_{\alpha4}|^2$. Therefore, the total width and lifetime of the HNLs also depend on the relevant mixing parameters. The lifetime dependence on the values of $|U_{\alpha4}|^2$ can have a huge impact on the position of the decay vertex of the HNL and hence on its possible signal at a detector. Setting small values for the $|U_{\alpha4}|^2$ means that the HNLs are being produced at a lower rate, but, at the same time, that these HNLs have a greater lifetime and therefore decay further away from the detector.\\

When we determine 
the individual partial widths of each channel, there is
a factor of two that differentiates between the decays of  Dirac and Majorana HNLs~\cite{Ballett:2019bgd}. For instance, a Dirac HNL can decay to charged pions only via $N\to e^-\pi^+$, while a Majorana one can also decay through $N\to e^+\pi^-$. This evidently has an effect on the rates of $\pi^+/\pi^-$ production from HNL decays but does not affect the partial decay widths. This means that CC mediated channels have the same partial widths for Dirac and Majorana neutrinos:
\begin{align}
\begin{split}
\Gamma(N_M\to l^- X^+)=\Gamma(N_D\to l^- X^+),\\
\Gamma(N_M\to l^+ X^-) = \Gamma(\bar{N}_D\to l^+ X^-).
\end{split}
\label{eq:relwidthsCC}
\end{align}
On the other hand, NC mediated channels do distinguish between Dirac and Majorana HNLs. This is because the contractions of the NC operator add an additional contribution to the differential decay width of the Majorana HNLs~\cite{Abada:2016plb,Ballett:2019bgd},
\begin{align}
d\Gamma(N_M\to \nu X) = d\Gamma(N_D\to \nu X) + d\Gamma(\bar{N}_D\to \bar{\nu} X).
\end{align}
Therefore, a factor of two appears when comparing the partial widths of NC mediated decays,
\begin{align}
\Gamma(N_M\to \nu X) = 2\Gamma(N_D\to \nu X).
\label{eq:relwidthsNC}
\end{align}
Equations (\ref{eq:relwidthsCC}) and (\ref{eq:relwidthsNC}) imply that the total widths $(\Gamma_T)$ of Majorana and Dirac HNLs are related by
\begin{align}
\Gamma_\text{T}(N_M)=2\Gamma_T(N_D),
\label{eq:totalDecayRate}
\end{align}
which translates into a difference between their lifetimes,
\begin{align}
\tau(N_M)=\frac{1}{2}\tau(N_D).
\label{eq:lifetime}
\end{align}
For very low masses $(m_N \ll m_e)$, the factor of two in Eq. (\ref{eq:relwidthsNC}) disappears  \cite{Kayser:1981nw}, making the total widths and lifetimes of Dirac and Majorana HNLs indistinguishable. Part of the mass range that we will explore in this work falls in the region of very low masses.

\begin{figure}[!h]
	\centering
	\includegraphics[width=.9\linewidth]{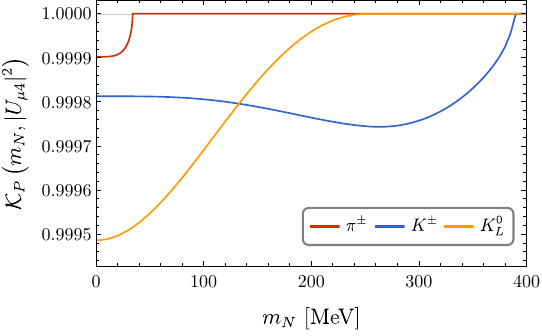}
  \caption{\label{img-br2SMLight}
    Suppression factor $\mathcal{K}^\mu\left(m_N, |U_{\mu4}|^2 = 10^{-4}\right)$ of muon neutrino production as a function of $m_N$.
  }
\end{figure}

At the end of this section, we will describe how the active neutrinos flux is affected by the production of HNLs. For this purpose, we will show how the SM parent meson's branching ratios are modified when the production of HNL occurs. Let us start by defining the SM total decay rate of the pion ($\Gamma^{\text{SM}}_\pi$):
\begin{align}
\Gamma^{\text{SM}}_\pi&=\Gamma^{\text{SM}}(\pi\rightarrow e\nu_e)+\Gamma^{\text{SM}}(\pi\rightarrow \mu\nu_\mu),
\end{align}
and the decay rate with heavy neutral leptons ($\Gamma^{\text{BSM}}_\pi$):
\begin{align}
\begin{split}
	\Gamma^{\text{BSM}}_\pi=&\ \Gamma^{\text{BSM}}(\pi\rightarrow e\nu_e)+\Gamma^{\text{BSM}}(\pi\rightarrow \mu\nu_\mu)
		+\Gamma(\pi\rightarrow NX)\\
		\approx&\ \Gamma^{\text{SM}}(\pi\rightarrow e\nu_e)\left(1-\frac{|U_{e4}|^2}{2}\right)\\
		&\ +\Gamma^{\text{SM}}(\pi\rightarrow \mu\nu_\mu)\left(1-\frac{|U_{\mu4}|^2}{2}\right)\\
		&\ +\Gamma(\pi\rightarrow NX).
\end{split}
\end{align}
The branching ratio of $\nu_\mu$ production from pion decays in the presence of HNLs can then be written as
\begin{align}
\begin{split}
	\text{BR}^{\text{BSM}}(\pi\rightarrow\mu\nu_\mu)
		&=\frac{\Gamma^{\text{BSM}}(\pi\rightarrow \mu\nu_\mu)}{\Gamma^{\text{BSM}}_\pi}\\
	&\approx\frac{\Gamma^{\text{SM}}(\pi\rightarrow \mu\nu_\mu) \left(1-\frac{|U_{\mu4}|^2}{2}\right)}{\Gamma^{\text{SM}}_\pi}
		\cdot \frac{\Gamma^{\text{SM}}_\pi}{\Gamma^{\text{BSM}}_\pi}\\
	&\approx\text{BR}^{\text{SM}}(\pi\rightarrow\mu\nu_\mu)
		\cdot \frac{\Gamma^{\text{SM}}_\pi}{\Gamma^{\text{BSM}}_\pi}\left(1-\frac{|U_{\mu4}|^2}{2}\right).
\label{pimuon}
\end{split}
\end{align}
A similar relation can be found for the branching ratio of $\nu_e$ production from pion decays:
\begin{align}
\begin{split}
\text{BR}^{\text{BSM}}(\pi\rightarrow e\nu_e) 
&\approx\text{BR}^{\text{SM}}(\pi\rightarrow e\nu_e)
		\cdot \frac{\Gamma^{\text{SM}}_\pi}{\Gamma^{\text{BSM}}_\pi}\left(1-\frac{|U_{e4}|^2}{2}\right),
\label{pielectron}
\end{split}
\end{align}
where $\text{BR}^{\text{SM}}(\pi\rightarrow\mu(e)\nu_{\mu(e)})$
represents the branching ratio of $\nu_\mu(\nu_e)$ production from pion decays in the SM. We can see that the introduction of HNLs causes the production of either muon or electron neutrinos from pions to be suppressed by the factor
\begin{equation}
\mathcal{K}^\alpha_\pi\left(m_N, |U_{\alpha4}|^2\right)= 
\displaystyle\frac{\Gamma^{\text{SM}}_\pi}{\Gamma^{\text{BSM}}_\pi}\left(1-\frac{|U_{\alpha 4}|^2}{2}\right),
\label{eq:suppresion}
\end{equation}
with $\alpha=e,\mu$. Fig. \ref{img-br2SMLight} illustrates the dependence on $m_N$ of the factor $\mathcal{K}^\mu$ for several parents assuming $|U_{\mu4}|^2 = 10^{-4}$. For each meson, the suppression factor acts only up to a maximum HNL mass due to kinematical constraints, which are the same constraints shown in Table \ref{tab:prodchannels} and Fig. \ref{fig:br2hnlLight}. Although the effect is small, the high luminosity of DUNE makes it possible to use this effect to set limits on the heavy neutral leptons parameters.\\

Thus, each particle capable of producing active neutrinos can now produce HNLs, leading to a suppression of active neutrinos at DUNE. The latter happens for all neutrino flavours even when only one mixing $|U_{\alpha4}|^2$ is turned on. In fact, we can see from Eqs.~(\ref{pimuon}) and (\ref{pielectron}) that, if we set $|U_{\alpha4}|^2=0$, the production of the active neutrinos $\nu_\alpha$ is still suppressed by the factor $\Gamma^{\text{SM}}_\pi/\Gamma^{\text{BSM}}_\pi$. As we will show further ahead, the reduction in the active neutrino flux would imply the possibility that they do not reach the DUNEND, decreasing the number of expected CC events at this facility.


\section{Experimental Setup}

In order to simulate how the presence of HNLs affects the number of $\nu$ CC events at DUNE, we based our experimental setup in the DUNE Near Detector, described in Ref. \cite{DUNE:2020ypp}.

\begin{figure}[!h]
	\centering
	\includegraphics[width=.9\linewidth]{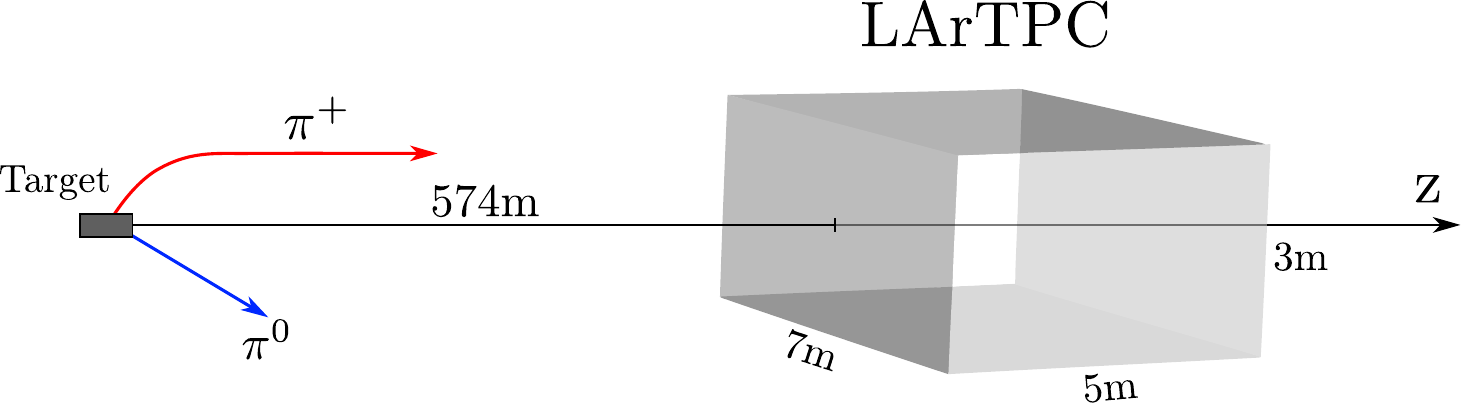}
  \caption{\label{fig:expSetup}
    Experimental setup for the LArTPC in neutrino mode (not to scale). Charged particles are deflected by the magnetic horns.
  }
\end{figure}

We assume that the LBNF-DUNE beam collides protons with 120 GeV of energy into a graphite target, producing $1.47\times10^{21}$ POTs per year. At each collision, several mesons are produced, including mostly pions, kaons and charmed mesons.

The muons and long-lived charged mesons ($\pi^\pm$ and $K^\pm$) produced are deflected by focusing magnetic horns located right after the target; as a consequence, their trajectories end up preferably oriented along the beam axis, as shown schematically in Fig. \ref{fig:expSetup}. On the other hand, the trajectories of neutral mesons ($D^0$, $K_L^0$ and $\pi^0$), tau leptons and short-lived charged heavy mesons ($D^\pm$ and $D_s^\pm$) are not affected by the focusing horns. Most particles decay in flight inside the decay pipe, a cylinder with a length of 230 m and a diameter of 2 m; however, a small number of long-lived particles reach the end of the decay pipe and decay at rest at the decay pipe's surface.

The Near Detector Liquid Argon Time Projection Chamber (LArTPC) is located at 574 m from the target. It has the shape of a parallelepiped with width and height (both transverse to the beam direction) of 7 m and 3 m, respectively, and a length of 5 m in the beam direction. The LArTPC is filled with a fiducial mass of 50 tons of liquid Argon. There is also the Multi-Purpose Detector (MPD), which is a magnetic spectrometer designed to study particles exiting the LArTPC that contains a one-ton high-pressure cylindrical gaseous argon time projection chamber. Since we are interested in the effects of HNLs on the $\nu$ CC events at the DUNE Near Detector, we will not take into account the MPD in our simulation setup because its impact on our results is negligible.  

We take also into account the possibility of moving the detectors to several off-axis positions along the x-axis, a setup known as DUNE-PRISM \cite{DUNE:2021tad}.


\section{SIMULATION ROUTE FOR HNLS}
\subsection{Parents Production}

\begin{figure}[!h]
	\centering
	\includegraphics[width=.9\linewidth]{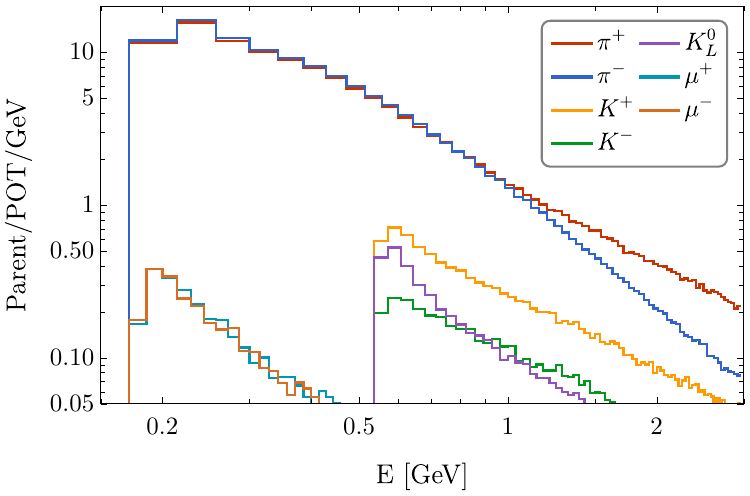}
  \caption{\label{fig:AllParents}
    Spectra of light particles capable of producing HNLs in the DUNE beam. Different bin widths have been used for different particles.
  }
\end{figure}

For the simulation of the production of HNLs from light mesons, we used the data provided by the DUNE Beam Interface Working Group (BIWG) \cite{biwg}, which makes use of GEANT4 \cite{GEANT4:2002zbu,Allison:2016lfl} and FLUKA \cite{Ferrari:2005zk,Bohlen:2014buj}. This data includes information about the decay positions and momenta of pions, kaons and muons after they exit the focusing horns. The most abundant light parent in DUNE is the pion, followed by kaons and finally muons, as can be seen in Fig. \ref{fig:AllParents}. In this work we will consider that the neutrino CC event rates might have an overall normalization uncertainty of up to 20\% due to uncertainties in the modeling of production of mesons and leptons at the DUNE target and neutrino cross sections. We encapsulate this uncertainty by a parameter $\sigma_a$ that varies from 0.05 to 0.2. Setting $\sigma_a=0.05$ is equivalent to assume a 5\% overall normalization uncertainty in the DUNE neutrino CC event rates, whereas $\sigma_a$ represents an uncertainty of 20\%.

The production of HNLs from heavier particles such as $D$ mesons and $\tau$ leptons is also possible, but it is expected to have a negligible effect on the active neutrino flux, which is totally dominated by production from lighter mesons. In order to test the relevance of HNL production from these heavy particles, we used PYTHIA8 \cite{Sjostrand:2014zea} to estimate the neutrino flux generated by $D^0, \bar{D}^0, D^\pm, D_s^\pm$ and $\tau^\pm$ at DUNE. We observed that these heavy parents do not contribute significantly to the DUNE neutrino flux and hence the production of HNLs coming from them will have a negligible effect on the number of CC events. Consequently, our analysis is restricted only to the production of HNLs from light mesons and muons. 

\subsection{Production of HNLs }
\label{sec:pythia8}

The production and decay chain of a HNL will depend on its mass, the mass of its parent, the nature of its parent (lepton, scalar meson or vector meson), the parent decay channel, the HNL nature (Dirac or Majorana), the HNL decay channel and the value of the mixing parameter involved. In principle, we could turn on, simultaneously, the three mixing parameters $|U_{\alpha 4}|^2$, $\alpha=e,\mu$ and $\tau$; however, in our analysis we will consider only one non-zero mixing parameter at a time. 

Given a HNL mass and nature, we gave PYTHIA8 the kinematic information of the parents and let it handle the kinematics of all the HNL production and decay chain, up to final active neutrinos. As expected, the HNL production and decay channels are weighted with their corresponding branching ratios. 

\begin{figure}[!h]
	\centering
	\includegraphics[width=.9\linewidth]{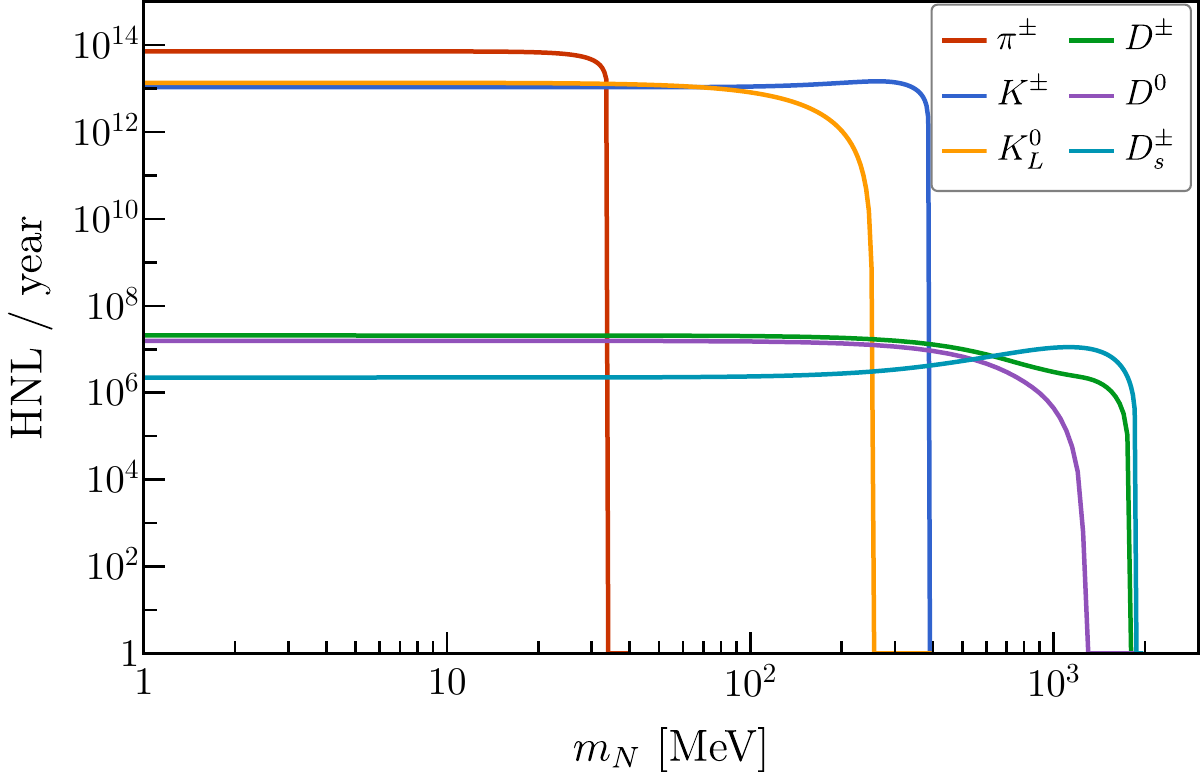}
  \caption{\label{fig:br2hnl}
    Heavy Neutral leptons produced from mesons in one year in neutrino mode for $|U_{\mu4}|^2=10^{-4}$.
  }
\end{figure}

In Fig. \ref{fig:br2hnl} we show the number of HNLs produced at DUNE from mesons decays in one year and in neutrino mode for $|U_{\mu4}|^2=10^{-4}$.  Production from pion decays dominates at low masses, followed by charged and neutral kaons. The spectrum endpoint for pions and kaons corresponds to the maximum allowed $m_N$ displayed in Table~\ref{tab:prodchannels} when they decay into muons. For completeness, we also present the production from charmed mesons, which, as expected, is comparatively smaller and completely overshadowed for masses below 387.81 MeV. Above this threshold, HNL production from pions and kaons is kinematically forbidden, and the contribution from 
charmed meson decays dominates. This contribution is several orders of magnitude smaller than the one from light mesons, as we already claimed.

\subsection{Decay of HNL - Active Neutrinos }
\label{sec:hnlnuactv}

We focus on the active neutrinos produced from the HNL decays. We are interested in differentiating the number of these neutrinos that fall within the detector's geometrical acceptance from those outside of it. With this aim, we parametrize the probability that an active neutrino hits the detector by two distances along the HNL propagation axis. These distances represent two different decay vertices of the HNL and are calculated considering the geometrical coverage of the detector and the kinematical information provided by PYTHIA8, which depends on its lifetime, production vertex, velocity and the direction of propagation of the active neutrino. The aforementioned probability is given by:
\begin{align}
w(d_1,d_2)=\exp\left(-\frac{d_1}{v \gamma \tau_0}\right)-\exp\left(-\frac{d_2}{v \gamma \tau_0}\right),
\label{eq:probdet}
\end{align}
where $v$ is the HNL's velocity, $\gamma$ its Lorentz factor and $\tau_0$ its proper lifetime.

\begin{figure}[!h]
	\centering
	\includegraphics[width=.9\linewidth]{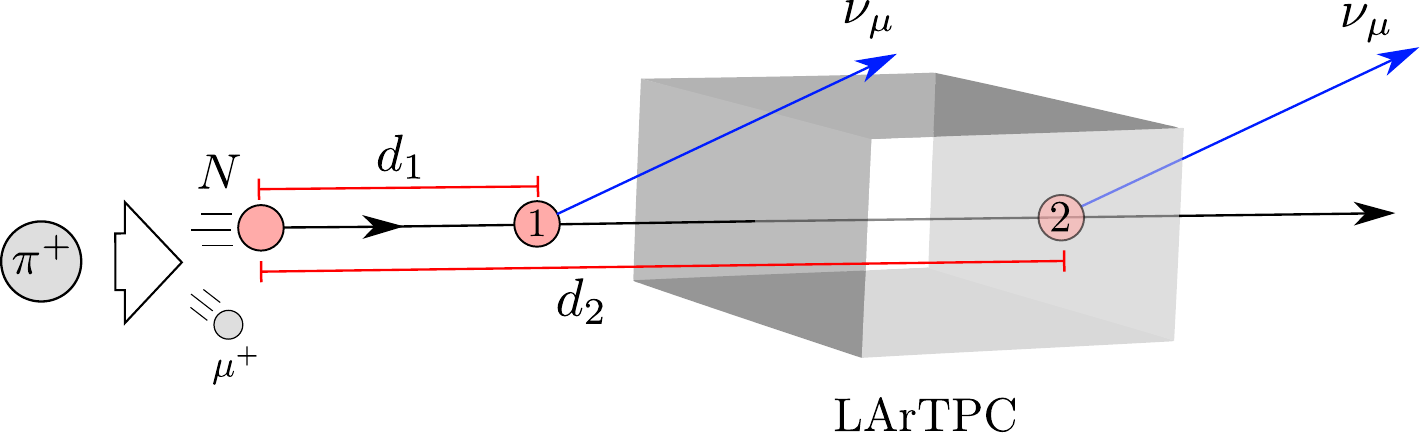}
  \caption{\label{fig:wg_nu}
    A HNL $N$ propagates and decays into an active neutrino $\nu_\mu$. If the HNL decays between positions 1 and 2, the active neutrino $\nu_\mu$ hits the LArTPC.
  }
\end{figure}

For illustrative purposes, we present in Fig. \ref{fig:wg_nu} the scheme of the explained above, for the case when the HNL moves along the beam axis. It is clear that our analysis is general and takes into account the tridimensional shape of the LArTPC and all the possible ways in which an active neutrino might enter the detector, including cases where the HNL is outside the detector coverage.

It is important to mention that when we deactivate the HNL production, 
we reproduce the (pure SM) active neutrino fluxes arriving at the LArTPC predicted by the DUNE Collaboration \cite{DUNE:2021cuw}. 

\begin{figure}[!h]
	\centering
	\includegraphics[width=.9\linewidth]{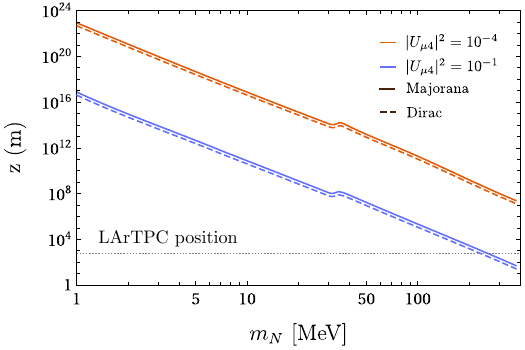}
  \caption{\label{fig:zdec}
    Average HNL's decay positions proyected along the Z axis for $|U_{\mu4}|^2=10^{-4}$ and $|U_{\mu4}|^2=10^{-1}$. The dotted line represents the position of the LArTPC.
  }
\end{figure}

In Fig. \ref{fig:zdec} we display the average HNLs' decay positions measured from the target and projected along the Z-axis for $|U_{\mu4}|^2=10^{-4}$ and $|U_{\mu4}|^2=10^{-1}$ and for Dirac and Majorana HNLs. The dotted line represents the position of the LArTPC, which is located at $z=574\text{ m}$. Given that the lifetime of the HNL is inversely proportional to $|U_{\mu4}|^2$, we can see that, as long as the mixing decreases, the average decay positions at Z increase. In the mass range we studied, for $|U_{\mu4}|^2=10^{-4}$, on average, all the HNLs decay behind the LArTPC; hence, one active neutrino is lost in the DUNE flux at the LArTPC per each HNL produced. On the other hand, for $|U_{\mu4}|^2=10^{-1}$, the average HNL decay position coincides with the LArTPC location at $m_N\approx 255$ MeV, which implies that, above this mass, the HNLs decay mainly before the detector.

We also note that in both cases there is a small increase in the average decay positions around 30 MeV. This happens because the production of HNLs from pion decays becomes kinematically forbidden around this energy and decays from kaons start to dominate. This makes the average HNL more energetic and therefore it can travel larger distances before decaying.

\subsection{Oscillation effects in Active Neutrinos from meson decays}
\label{sec:hnlnuactv}

The existence of HNLs forces us to modify the neutrino oscillation probabilities. Therefore, the effects of neutrino oscillations have to be taken into account in our simulations. Particularly, the place where neutrino oscillations can affect our results is in the disappearance of active neutrinos produced in meson decays. The survival probability of these active neutrinos is given by
\begin{align}
	 P_{\nu_\alpha\rightarrow\nu_\alpha} = & -4\left(1-|U_{\alpha4}|^2\right)|U_{\alpha4}|^2\sin^2\left(\frac{1.27m_N^2L}{E}\right)e^{-\frac{\Gamma_4L}{2}} \nonumber\\
	 &+2\left(1-|U_{\alpha4}|^2\right)|U_{\alpha4}|^2e^{-\frac{\Gamma_4L}{2}}\nonumber\\
	 &+\left(1-|U_{\alpha4}|^2\right)^2 + |U_{\alpha4}|^4e^{-\Gamma_4L}, \label{eq:pinv}
\end{align}
where $E$ represents the energy of the active neutrino, $L$ the distance that it travels before reaching the DUNEND, $\Gamma_4$ the decay rate of the HNL and we have considered that the mass of the active neutrino is negligible when compared to the HNL mass $m_N$. This survival probability will effectively decrease the number of active neutrinos that reach the DUNE ND and the number of neutrino CC events at the Near Detector Complex. For completeness, we incorporated Eq. (\ref{eq:pinv}) in our simulations as an extra weight for each active neutrino. \\

There is also the possibility of oscillation of HNLs into active neutrinos. However, since the HNL flux is very small when compared to the active neutrino flux, the effects of these oscillations in the neutrino CC event rates are negligible and were not considered in this work.

\section{results}

\subsection{Impact on CC events at DUNEND}

As we can infer from what we have shown before, the DUNE neutrino flux fired at the DUNEND will be affected by the production of HNLs. Each HNL produced from the decay of its parent meson (or muon) replaces one active neutrino in the SM DUNE neutrino flux. In principle, there is a possibility to recover this active neutrino since the HNL can decay into one or more active ones, which, depending on their direction, could or not impact the DUNEND. However, as it is demonstrated in Fig. \ref{fig:zdec}, it is unlikely that a relevant portion of these spurious active neutrinos would be created before or inside the LArTPC of the DUNEND for the mass range used in this work. This decrease in active neutrinos translates into a decrease in the CC event rates at the LArTPC. Our strategy is to use this deficit of CC events as an indirect signal of the existence(production) of HNLs at the DUNE neutrino flux. Hence, in that sense, we are conducting an indirect search for HNLs. This indirect method for searching for HNLs is complementary to the direct searches~\cite{Berryman:2019dme}, which look for HNL decays inside one of the DUNE's detectors. As we will show in the following sections, our method can work comparatively better than direct searches for masses below 10 MeV and is sensitive to masses below 1 MeV, a region primarily inaccessible through direct searches.

The deficit in the total CC event rates depends on the mass of the HNL, the value of $|U_{\alpha4}|^2$ and the off-axis position of the detector. In order to have a first estimate of the maximum significance of this deficit allowed by current limits on the mixing parameters, we calculated the active neutrino flux in presence of HNLs using the maximum values of $|U_{\alpha4}|^2$ allowed by accelerator experiments at 90\% confidence level \cite{Bolton:2019pcu} and then convoluted these fluxes with GENIE 2.8.4 \cite{Andreopoulos:2009rq} CC  inclusive cross sections.\\

\begin{figure}[!h] 
	\centering
	\includegraphics[width=.9\linewidth]{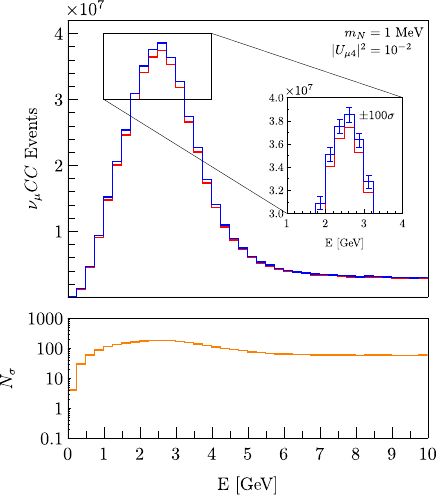}
  \caption{\label{fig:spectrum1mev}
    $\nu_\mu$ CC event rates for $m_{N}=1\text{ MeV}$ assuming the maximum value allowed for $|U_{\mu4}|^2$ at 90\% confidence level, on-axis position, 10 years of operation and $\sigma_a=0$. The error bars are amplified by 100.
  }
\end{figure}

In order to get an idea of the significance of the change in the neutrino CC event rates, we will ignore all systematic uncertainties in the neutrino flux prediction and work in the ideal case of no systematic uncertainties $\sigma_a=0$. Figure \ref{fig:spectrum1mev} shows the $\nu_\mu$ CC event rates at the LArTPC for $m_{N}=1\text{ MeV}$ and $|U_{\mu4}|^2=10^{-2}$ assuming Majorana neutrinos, on-axis position, 10 years of operation (5 in neutrino and 5 in antineutrino mode) and $\sigma_a=0$. The significance of the change in the number of the CC events in each bin is estimated by
\begin{align}
N_\sigma = \frac{|N^{\text{BSM}}-N^{\text{SM}}|}{\sqrt{N^{\text{SM}}}}=\frac{|\Delta N|}{\sigma}
\end{align}
where $N^\text{SM}$ represents the expected number of CC events assuming only SM interactions and $N^\text{BSM}$ the number of CC events when HNLs are produced. As we mentioned before, we are also ignoring all normalization uncertainties in the CC event rates, so that $\sigma=\sqrt{N^\text{SM}}$ is the uncertainty in each bin. Due to the high luminosity of the DUNE experiment, under this setup, the production of HNLs causes a decrease in the total number of CC event rates on the order of $10^6$ events near $2.5 \text{ GeV}$. This implies a deviation from the SM prediction by approximately 100$\sigma$ around this energy. This indicates that DUNE's sensitivity to $|U_{\mu4}|^2$ might be beyond the current experimental limits for this particular HNL mass.\\ 

\begin{figure}[!h]
	\centering
	\includegraphics[width=.9\linewidth]{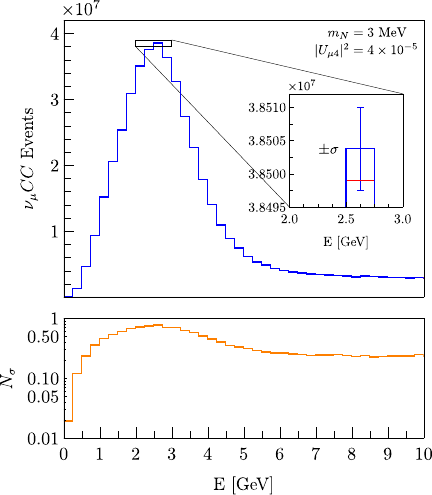}
  \caption{\label{fig:spectrum30mev}
    $\nu_\mu$ CC event rates for $m_{N}=3\text{ MeV}$ assuming the maximum value allowed for $|U_{\mu4}|^2$ at 90\% confidence level, on-axis position, 10 years of operation and $\sigma_a=0$.
  }
\end{figure}

As the HNL mass increases, its production is suppressed, and, consequently, its presence on the active neutrino flux is reduced. As an example of the latter, we display in Fig. \ref{fig:spectrum30mev} the event rates for $m_N=3\text{ MeV}$ and the maximum value allowed for $|U_{\alpha4}|^2$ by experiments at 90\% confidence level for this mass. In this case, there is a (small) deviation, from the SM prediction, lower than 1$\sigma$. This happens because of the tighter constraint on the mixing parameter.\\

We have shown that in the ideal case of no systematic uncertainties $\sigma_a=0$ DUNE will have good sensitivity for indirect hints of the existence of low mass HNLs, which are  evidenced by a decrease in the neutrino CC event rates at the LArTPC. Of course, once systematic uncertainties are considered, the sensitivity and the limites are expeted to decrease considerably.


\subsection{Sensitivity }

In order to estimate the future sensitivity of DUNE to HNLs due to the deficit of neutrino CC events, we have to consider that the predictions of our simulations carry systematic uncertainties related to the distributions of hadron production at the DUNE target, the neutrino CC cross section uncertainties, among others. We will incorporate these uncertainties in our calculations by assuming an overall normalization uncertainty in the spectra, which, in practice, means that the values of the event rates are not completely known and can fluctuate by a certain amount. This overall normalization uncertainty will be represented by the the parameter $\sigma_a$ that takes the values $\sigma_a =$ 0.05, 0.1 and 0.2, which are equivalent to overall normalization uncertainties of 5\%, 10\% and 20\%, respectively. We are also considering shape uncertainties in each bin that are represented by the parameters $\sigma_{ai}$; for simplicity, we consider that $\sigma_{ai}=\sigma_a$ for all bins. We estimate the sensitivity of DUNE to $(m_N,|U_{\alpha4}|^2)$ through the following $\chi^2$ \cite{Fogli:2002pt,Rodejohhan:2018zcz}:
\begin{align}
\chi^2 = \frac{a^2}{\sigma_a^2}  + & \sum_{{\nu}_{e},{\nu}_{\mu},{\bar{\nu}}_{e},{\bar{\nu}}_{\mu}}\left[ \sum_{i=1}^{\text{nbin}}\frac{a_i^2}{\sigma_{ai}^2} + \sum_{i=1}^{\text{nbin}}\frac{\left(N_i^{\text{SM}}-N_i^{\text{BSM}}(1+a+a_i)\right)^2}{N_i^\text{SM}} \right],
\label{eq:chi2}
\end{align}
where $N_i^{\text{BSM}}$ represents the neutrino CC events in the i-th bin when HNLs are produced and $N_i^{\text{SM}}$ the DUNE prediction of CC events in i-th bin according to the Standard Model. The nuisance parameters $a$ and $a_i$ encompass the normalization uncertainties and allow for the values of $N_i^\text{BSM}$ to fluctuate; these parameters are always profiled in the calculation of the $\chi^2$. We must note that the fact of combining all the neutrino flavours in our definition of $\chi^2$ is fundamental for improving the sensitivity of our results. The tau neutrinos are not considered since their contributions to the $\chi^2$ are negligible.
\\

The deficit of neutrino CC events at DUNE is an indirect signal of HNLs. Therefore, in the case that no significant deficit is found, the absence of this deficit can be used to set limits on the values of the parameters $(m_N,|U_{\alpha4}|^2)$ with a particular confidence level. We calculated the value of $\chi^2$ in the parameter space $10^{-6}\text{ MeV}<m_N<10^7\text{ MeV}$ and $10^{-12}<|U_{\alpha^4}|^2<1$ for $\alpha = e, \mu$ and $\sigma_a$ = 0.5, 0.1, 0.2 and then used these values to estimate the limits that DUNE might be able to set to the parametes $(m_N,|U_{\alpha4}|^2)$ at 90\% confidence level.

\begin{figure*}
\centering
\includegraphics[width=\linewidth]{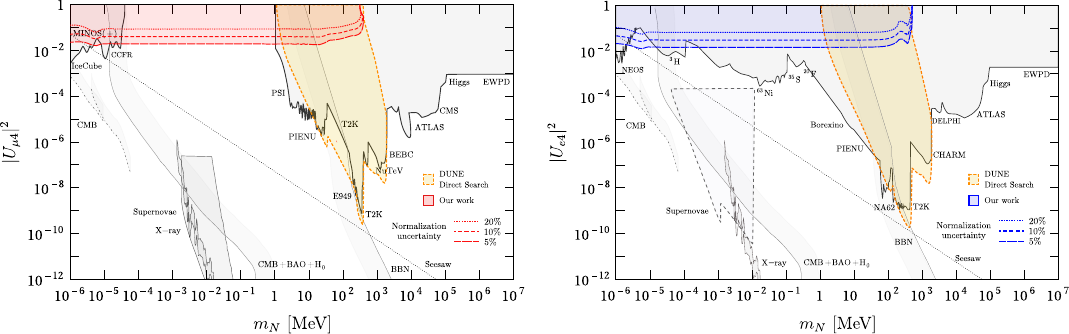}
\caption{\label{fig:limits-01}
	Estimated limits of DUNE to $|U_{\mu4}|^2$ (left, red) and $|U_{e4}|^2$ (right, blue) at 90\% confidence level by CC events disappearance at the LArTPC of the DUNEND, for 10 years of operation (5 in neutrino and 5 in antineutrino mode) and on-axis position. The regions of experimental constraints (gray) were taken from \cite{Bryman:2019ssi,Arguelles:2021dqn,Bolton:2019pcu}. The estimated sensitivity of DUNE obtained in \cite{Berryman:2019dme} by direct searches of HNL decays is shown for comparison.
}
\end{figure*}

Our results are presented in Fig. \ref{fig:limits-01}. The left panel of this figure shows the estimated DUNE sensitivity to $|U_{\mu4}|^2$ at 90\% confidence level on the LArTPC assuming Majorana neutrinos, ten years of operation (five in neutrino and five in antineutrino mode) and on-axis position. In our analysis, the CC event rates from all neutrino flavors are considered (read the discussion at the end of section~\ref{theory}). For masses close to 1 eV, the limits decrease because, for the typical energies and flight distances of active neutrinos at DUNEND, the probability of neutrino oscillations into HNLs tends to zero as the value of $m_N$ approaches 1 eV. Right above 1 MeV, the limits start to oscillate since the survival probability of the active neutrinos is sensitive to $m_N$. For masses between 10 eV and 10 MeV, the limits are independent of $m_N$. The latter is because of three factors. The first one is the averaging out of the neutrino oscillations into HNLs for large values of $m_N$. The second one is that, for these very low masses, the total number of HNLs produced is practically independent of $m_N$ (see Fig. \ref{fig:br2hnl}). The other factor is that the HNL lifetime for lower masses is enormous (see Fig.\ref{fig:zdec}), decaying all of them far away from the detector without the possibility of leaving a trace on it. As we already know, above $m=33.91\text{ MeV}$, the production channel $\pi^+ \rightarrow \mu^+N$ is kinematically forbidden, and there is a sudden loss in the sensitivity. As the mass increases, production from charged kaons starts to dominate and does so up to the end of the curve, which is at 387.81 MeV. For instance, for $\sigma_a = 0.05$ and $\sigma_a = 0.2$, the sensitivity of DUNE below 10 MeV is around $|U_{\mu4}|^2<2\times10^{-2}$ and $|U_{\mu4}|^2<8.5\times10^{-2}$, respectively. We point out that even in the conservative case of $\sigma_a = 0.2$ our limits are competitive with direct searches below 1.3 MeV.

The right panel of Fig. \ref{fig:limits-01} shows the expected DUNE sensitivity when we turn on $|U_{e4}|^2$ being the other ones zero. The rest of the characteristics are the same as for the left panel. In general, the sensitivity pattern is similar to the one observed for the left panel. The limits oscillate close to 1 eV and for higher masses they become mass independent since most HNLs decay behind the LArTPC. Above 10 MeV, the pion decay channel $\pi^\pm\rightarrow e^\pm N$ starts to dominate because, in contrast to $\pi^\pm\rightarrow e^\pm\accentset{\brobor}{\nu_e}$, it is less suppressed by helicity due to the larger size of the HNL mass. This effect decreases the number of both $\nu_e$ and $\nu_\mu$ CC events according to the suppression factor in Eq. (\ref{eq:suppresion}), affecting the CC event rates of both eletron and muon neutrinos. At around 139 MeV, HNL production from pion decays becomes kinematically forbidden, which translates into a decrease in the sensitivity. Finally, the curve ends when production from kaons is kinematically forbidden at 493.17 MeV. For $\sigma_a = 0.05$ and $\sigma_a = 0.2$, the sensitivity of DUNE below 10 MeV is around $|U_{e4}|^2<1.5\times10^{-2}$ and $|U_{e4}|^2<6.5\times10^{-2}$, respectively. Even in the conservative case of $\sigma_a = 0.2$, our limits are competitive with direct searches below 1.3 MeV and also provide a small increase of sensitivity by a factor of 1.5 around 5 eV in comparison with experimental constraints.\\

Although we are making our calculations for ten years of exposure, it is important to point out that our sensitivity for $|U|^2$ increases only slightly when compared with one year of exposure. If we had not included systematic uncertainties, the limits would roughly improve as $ \frac{|U|^2}{\sqrt{T}}$, where $T$ represents the exposure time; in this ideal scenario, after 10 years of operation, the limits would improve by a factor of around $1/\sqrt{10}\approx0.32$. However, introducing uncertainties in our $\chi^2$ prescription heavily penalize the sensitivity of our approach: in this more realistic scenario, after 10 years of operation, the limits improve by only a factor of around 0.9 in comparison with one year of exposure. Therefore, in the context of our analysis, the first year of operation of DUNE is the most important.

\begin{figure}[!h]
	\centering
	\includegraphics[width=.9\linewidth]{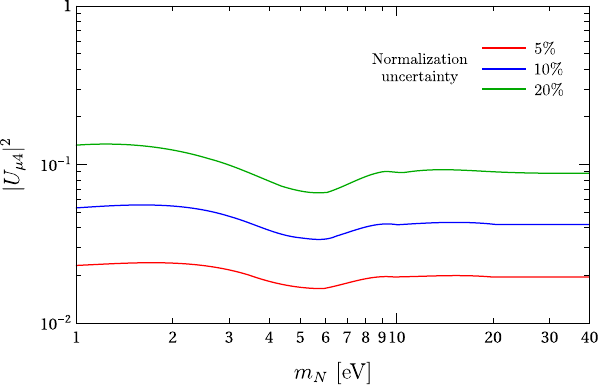}
  \caption{\label{fig:wiggle}
    Zoom of the sensitivy to $|U_{\mu4}|^2$ of Fig. \ref{fig:limits-01}. The oscillation of the sensitivity near 1 eV is produced by the oscillation effects of Eq. (\ref{eq:pinv}).
  }
\end{figure}

Another important remark must be done about the effects of neutrino oscillations in this work. Neutrino oscillations involving HNLs are only relevant when $m_N\sim \text{1 eV}$. Since our analysis starts a 1 eV, the effects of neutrino oscillations will only be visible as a wiggle at the beginning of our sensitivity plots. For completeness, in Fig. \ref{fig:wiggle} we show a zoom of the left plot of Fig. \ref{fig:limits-01}. We can see that close to 1 eV the sensitivy oscillates as expected, but this effect is small and only restricted to the low tail of our sensitivity plot.

We must point out that our results are blind to the Dirac or Majorana nature of the HNL. The distinction between Dirac and Majorana HNLs is usually performed in direct searches by analyzing the distributions of charged mesons and leptons produced when the HNL decays inside the detector. We are not looking into the direct search mode since it has already been discussed in \cite{Berryman:2019dme}. Besides their decay products, Dirac and Majorana HNLs can also be differentiated by their lifetimes due to the factor of two present in Eq. (\ref{eq:relwidthsNC}). However, this effect is not relevant for us because, for the mass range we studied and small mixings, almost all the HNL decays occur behind the LArTPC, as shown in Fig. \ref{fig:zdec}. Furthermore, as we have discussed in section~\ref{theory}, for very low $m_N$ the Dirac and Majorana neutrinos are indistinguishable. Thus, we can conclude that nearly all the active neutrinos produced from the HNL decays are lost independently of the nature of neutrinos. In this way, the critical magnitude in our analysis is the production rate of HNLs, which is independent of the nature of neutrinos, so the deficit of the CC event rates is independent too. Therefore, it would not be possible to distinguish between Dirac or Majorana neutrinos through the approach presented here.

\subsection{Off-axis sensitivity} 

\begin{figure}[!h]
	\centering
	\includegraphics[width=.9\linewidth]{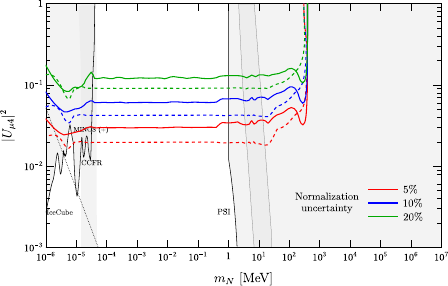}
  \caption{\label{fig:sensi1111}
    Comparison between on-axis (dashed) and 30 m off-axis (solid) estimated sensitivities of DUNE to $|U_{\mu4}|^2$ at 90\% confidence by neutrino CC events disappearance for 10 years of operation (5 in neutrino and 5 in antineutrino mode). The regions of experimental constraints were taken from \cite{Bryman:2019ssi,Arguelles:2021dqn,Bolton:2019pcu}.
  }
\end{figure}

The DUNE experiment also considers the possibility of moving the DUNE near detectors horizontally, a setup known as DUNE PRISM. We move the LArTPC by up to 30 m horizontally while maintaining the rest of the simulation parameters and study the impact in our estimated sensitivities. The results are shown in Fig. \ref{fig:sensi1111}, where the all the lines represent the sensitivities at 90\% confidence level and the dashed curves represent the on-axis sensitivities. We see that the effect of moving the detector to an off-axis position does not affect considerably the limits, although the curves are less smooth due to the decrease in statistics. However, we see that the sensitivitie increases at off-axis positions for masses close to 100 MeV.

\subsection{Allowed regions for $(m_N,|U_{\alpha4}|^2)$}
\begin{figure}[!h] 
	\centering
	\includegraphics[width=.9\linewidth]{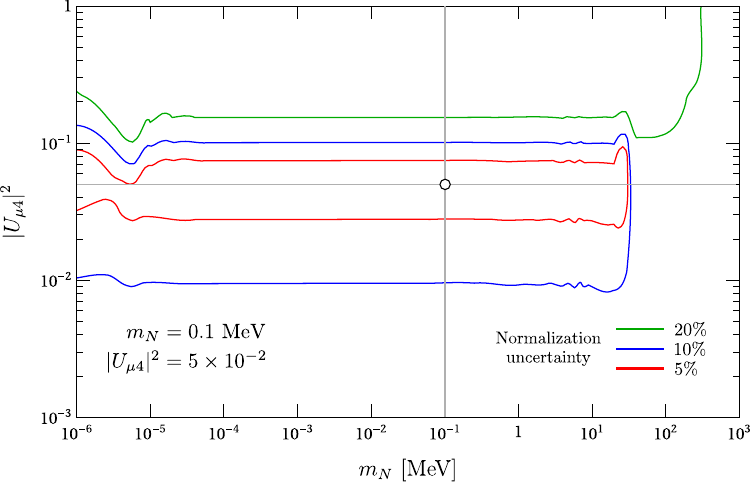} 
  \caption{\label{fig:figMixAB}
    95\% confidence level regions for $m_N=0.1\text{ MeV}$, $|U_{\mu4}|^2=5\times10^{-2}$, 10 years of operation (5 in neutrino and 5 in antineutrino mode), on axis position and several values of $\sigma_a$.  }
\end{figure}

We also explore the potential to constraint the $(m_N,|U_{\alpha4}|^2)$ parameter space region in the context of this indirect search. So, assuming that the disappearance CC events are originated by the presence of HNLs within the neutrino beam, we perform a $\chi^2$ analysis fixing our simulation in certain values of  $(m_N,|U_{\alpha4}|^2)$. The 95\% confidence regions for $m_N=0.1\text{ MeV}$ and $|U_{\mu4}|^2=5\times10^{-2}$ are presented in Fig. \ref{fig:figMixAB} for $\sigma_a = 0.05$ (red), $\sigma_a = 0.1$ (blue) and $\sigma_a = 0.2$ (green). For the case of 20\% uncertainty, the analysis only gives upper limits to the values of $|U_{\mu4}|^2$. For 5\% and 10\% uncertainties, the regions are bounded to the right, but extend to the left up to $m_N = 1$ eV, a mass degeneracy that reflects the fact that our approach is not sensitive to $m_N$ for low masses. For the case $\sigma_a = 0.05$ (red) the 95\% confidence region is sufficiently small that it is possible to constraint $|U_{\mu4}|^2$ within an uncertainty of 50\%. However, when we include larger systematic uncertainties such as $\sigma_a=0.1$ (blue) we find that we can only constraint the value of $|U_{\mu4}|^2$ within an uncertainty of 100\%. 

\section{Conclusions}
The cornerstone of this work is the analysis of other ways in which the active neutrino flux at DUNE is affected by the production of HNLs besides neutrino oscillations. We found that, in the presence of HLNs, the production rates of active neutrinos decrease and their angular distributions widen, which translates into a decrease of the number of $\nu_\mu$ and $\nu_e$ CC events in the LArTPC of the DUNEND. This neutrino disappeareance represents and indirect signal of HNLs at DUNE that is not due to neutrino oscillations, but rather to the kinematics of the meson and HNL decays. When combined with the effects of neutrino oscillations, it is possible to use this deficit in CC event rates to estimate limits to $|U_{e4}|^2$ and $|U_{\mu4}|^2$. We found that these limits are very sensitive to the uncertainty of the neutrino flux prediction at the DUNEND. In order to get conservative estimates of these limits, we considered  overall normalization uncertainties of up to 20\%. \\

For five years per mode (neutrino/antineutrino), on-axis configuration and a 5\% overall normalization uncertainty we get limits of $|U_{\mu4}|^2<2\times 10^{-2}$ and $|U_{e4}|^2<1.5\times 10^{-2}$ below 1.5 MeV. We also included a more pessimistic scenario of a 20\% systematic uncertainty and were still able to set bounds of $|U_{\mu4}|^2<8.5\times 10^{-2}$ and $|U_{e4}|^2<6.5\times 10^{-2}$ below 1.4 MeV. These limits are better than the ones predicted by DUNE direct searches or even placed in mass regions inaccessible to them. These bounds are still competitive for the off-axis configuration. Besides, we explore the capacity of determining the allowed parameter space region $(m_N,|U_{\alpha4}|^2)$ for the specific parameter values $m_N=0.1$ MeV and $|U_{\mu4}|^2=5\times10^{-2}$ and found that, although there is a large degeneracy in the value of $m_N$, it is possible to constraint $|U_{\mu4}|^2$ with uncertainties in the order of 50(100)\% for a 5(10)\% overall normalization uncertainty in the CC event rates. Finally, it is worth noting that the disappearance of CC events as a HNL signature is complementary to the direct observation or HNL decays, showing an attractive potential to be used in neutrino Near Detectors with high $\nu$ CC event rates.

\section*{Acknowledgments}
A. M. G. acknowledges funding by the Dirección de Gestión de la Investigación at PUCP, through grants No. DGI-2017-3-0019 and No. DGI 2019-3-0044. S. C. acknowledges CONCYTEC for the graduate fellowship under Grant No. 236-2015-FONDECYT. The authors also want to thank Zarko Pavlovic for useful remarks regarding the interpretation of the BIWG data, Margot Delgado de la Flor for her help at the initial stages of this work and R. E. Shrock, C. Arg\"uelles, I. Shoemaker, A. de Roeck, O. Peres and W. Rodejohann for their useful remarks on our first manuscript.

\bibliographystyle{unsrt}
\bibliography{biblio}

\end{document}